# Powder Spreading and Layer Deposition in Metal Powder Bed Fusion

*Antonello Astarita*
Dept. of Chemical, Materials and Production Engineering, University of Naples
"Federico II", Naples, Italy

**Abstract**
Powder spreading and layer deposition are fundamental stages of Powder Bed Fusion (PBF) technologies and play a critical role in determining process stability and final component quality. This chapter examines the mechanisms governing powder-bed formation, highlighting the interactions between powder characteristics, process parameters, and machine architecture. Particular attention is devoted to the influence of particle size distribution, morphology, cohesion, flowability, layer thickness, recoater velocity, and environmental conditions on powder-bed quality. The resulting powder-bed is discussed as a process state variable whose characteristics, including packing density, surface coverage, effective layer thickness, and spatial homogeneity, directly affect energy absorption, melt-pool stability, defect formation, and mechanical performance. The chapter also reviews the application of the Discrete Element Method (DEM) for modelling powder spreading phenomena and quantifying powder-bed quality metrics. Finally, the role of powder reuse, lifecycle management, and future developments involving process monitoring, digital twins, and data-driven optimization strategies is discussed, emphasizing the growing importance of powder engineering in advanced metal additive manufacturing.

## 1 Introduction

Powder Bed Fusion (PBF) technologies represent one of the most advanced and versatile families of metal additive manufacturing processes currently available. Their success is primarily attributable to the capability of producing components characterized by unprecedented geometric complexity, reduced material waste, and a high degree of design freedom. Unlike conventional subtractive or formative manufacturing routes, where the final geometry is obtained through material removal or deformation, PBF processes create components through the sequential deposition and consolidation of thin powder layers. This layer-by-layer paradigm constitutes the fundamental principle upon which the entire process is based. In Laser Powder Bed Fusion (LPBF) and Electron Beam Powder Bed Fusion (EB-PBF), the manufacturing cycle consists of a series of repeated operations. First, a thin layer of metallic powder is deposited onto the build platform. Subsequently, a focused energy source selectively melts or fuses specific regions according to the cross-sectional geometry of the component. Once the layer is consolidated, the build platform is lowered by a predefined distance and a new powder layer is deposited. This sequence is repeated hundreds or thousands of times until the complete component is manufactured. Although the energy source often receives most of the scientific attention, the actual material processed during each cycle is not the bulk alloy itself but rather the deposited powder layer. Consequently, the quality of the powder-bed directly determines the initial conditions under which melting occurs. Any variation in the characteristics of the deposited layer inevitably propagates through the subsequent stages of the process, affecting energy absorption, thermal transport, melt-pool evolution, solidification behavior, and final component properties. This observation highlights a fundamental difference between PBF technologies and many conventional manufacturing processes. In traditional manufacturing, the starting material is generally homogeneous and characterized by relatively stable properties. In contrast, each newly deposited powder layer in PBF represents a stochastic granular system whose characteristics continuously evolve throughout the build process. Therefore, understanding powder-bed formation becomes essential for achieving process stability and repeatability. For many years, research activities in additive manufacturing primarily focused on laser-material interaction, thermal phenomena, melt-pool dynamics, microstructural evolution, and mechanical performance. Within this framework, powder spreading was frequently regarded as a secondary operation whose sole purpose was to provide material for subsequent melting. Recent developments have significantly changed this perspective. Growing experimental evidence and numerical investigations have demonstrated that powder spreading should be considered a critical process stage rather than a simple preparatory step. The quality of the deposited powder-bed strongly influences the effectiveness of energy transfer and therefore affects virtually every aspect of component fabrication. The recognition of powder spreading as a key process variable has been driven by several observations. First, components manufactured using identical laser parameters often exhibit significant differences in density, surface quality, and mechanical properties. Second, many process defects traditionally attributed to laser settings have been shown to originate from powder-bed heterogeneities. Finally, numerical studies have demonstrated that variations in spreading conditions can substantially modify packing density, surface coverage, and effective layer thickness, thereby altering the response of the material during melting. As a consequence, modern process optimization strategies increasingly adopt a holistic perspective in which powder spreading, powder characteristics, melting conditions, and post-processing operations are considered interconnected elements of a single manufacturing chain. Within this framework, the powder-bed acts as the physical interface linking feedstock properties to final component quality.

## 2 Formation of the Powder Bed

The formation of a powder-bed is governed by the interaction between machine architecture, process parameters, and powder characteristics. Although different machine configurations exist, the fundamental principles remain largely similar across most industrial PBF systems. The spreading process begins with the presence of a powder reservoir containing the feedstock material. A recoating device, commonly referred to as a recoater, transfers powder from the reservoir to the build area. Depending on machine design, the recoater may consist of a rigid metallic blade, a flexible polymer blade, a rotating roller, or more sophisticated hybrid solutions. As the recoater moves across the build platform, powder particles accumulate in front of the spreading device, forming a dynamic powder heap. This heap continuously evolves as particles are transported, rearranged, and deposited onto the substrate. The interaction between the moving recoater and the granular material generates a complex flow regime controlled by gravity, friction, particle inertia, and cohesive interactions. The deposited layer is often described in terms of its nominal layer thickness, which corresponds to the vertical displacement imposed by the machine between successive layers. However, numerous studies have demonstrated that the actual thickness of the deposited layer may differ substantially from the nominal value. This discrepancy arises from powder compaction, particle rearrangement, substrate interactions, and local variations in powder distribution. The concept of effective layer thickness has therefore become increasingly important in additive manufacturing research. Unlike the nominal thickness specified by machine settings, the effective thickness represents the actual amount of powder deposited within a given region. Variations in effective layer thickness can significantly influence local melting conditions and contribute to the development of process-induced defects.

## 2.1 Characteristics of an ideal powder bed

From a manufacturing perspective, an ideal powder-bed should satisfy several requirements simultaneously. The first requirement is high packing density. A dense powder-bed contains a larger amount of material per unit volume and exhibits reduced void content. High packing density generally promotes improved thermal continuity, enhanced energy absorption, and more stable melting conditions. The second requirement is complete surface coverage. The deposited powder layer should fully cover the underlying substrate without leaving uncovered regions. Areas characterized by insufficient coverage may experience incomplete melting and subsequently develop lack-of-fusion defects. A third requirement concerns thickness uniformity. Variations in layer thickness generate local differences in the amount of material available for melting and may therefore alter thermal conditions during processing. Uniform layers contribute to process stability and reduce the probability of defect formation. Repeatability represents another key requirement. Since a typical LPBF component may contain thousands of layers, the spreading process must be capable of generating consistent powder-beds throughout the entire build. Small variations occurring during individual layers may accumulate over time and eventually affect component quality. Finally, the powder-bed should exhibit spatial homogeneity. Properties such as particle size distribution, packing density, and layer thickness should remain relatively constant throughout the build area. Spatial variability increases process uncertainty and complicates parameter optimization. These characteristics collectively define powder-bed quality and provide the basis for many of the metrics used to evaluate spreading performance.

## 2.2 Challenges in Powder Bed Formation

Despite its apparent simplicity, powder spreading is an inherently complex process characterized by significant stochasticity. One of the main challenges arises from the granular nature of the material itself. Unlike fluids or solids, granular systems exhibit behavior that combines aspects of both states. Particle motion depends on local interactions that continuously evolve during spreading, making the process highly sensitive to both powder properties and operating conditions. Particle segregation represents another important challenge. During spreading, particles of different sizes may separate due to differences in mobility and momentum. Larger particles frequently accumulate at the front of the powder heap, whereas smaller particles tend to occupy interstitial regions. This phenomenon may produce spatial variations in particle size distribution and packing density throughout the powder-bed. Cohesive interactions further complicate powder-bed formation. Since additive manufacturing powders typically exhibit particle diameters below 100 μm, cohesive forces become comparable to particle weight. These interactions promote particle clustering and may reduce powder flowability, particularly in highly reactive or moisture-sensitive materials. Layer-to-layer variability constitutes an additional source of uncertainty. The characteristics of a newly deposited layer are influenced not only by current spreading conditions but also by the morphology of the underlying surface. As the build progresses, thermal distortions, residual stresses, and geometric complexity may modify local spreading conditions, generating cumulative effects that affect process repeatability. The combination of these phenomena explains why powder-bed formation remains one of the most challenging aspects of Powder Bed Fusion technologies.

## 2.3 From Powder Bed Formation to Part Quality

The importance of powder spreading ultimately derives from its influence on final component performance. Powder-bed characteristics establish the initial conditions for energy absorption and therefore affect all subsequent stages of the manufacturing process. Packing density influences absorptivity and thermal conductivity. Surface coverage determines whether sufficient material is available for complete melting. Layer thickness affects energy distribution and melt-pool dimensions. Particle segregation alters local material distribution and may contribute to process variability. These effects directly impact melt-pool stability and defect formation. Poor powder-bed quality increases the probability of lack-of-fusion porosity, balling phenomena, surface roughness, and dimensional inaccuracies. Conversely, homogeneous and dense powder-beds promote stable melting conditions and improved densification. The relationship between powder-bed formation and component performance can therefore be summarized as a hierarchical chain:

Powder Characteristics → Powder Spreading → Powder-Bed Quality → Energy Absorption → Melt-Pool Stability → Defect Formation → Mechanical Properties.

The overall logic of the relationships discussed throughout this chapter is schematically summarized in Fig. 1.

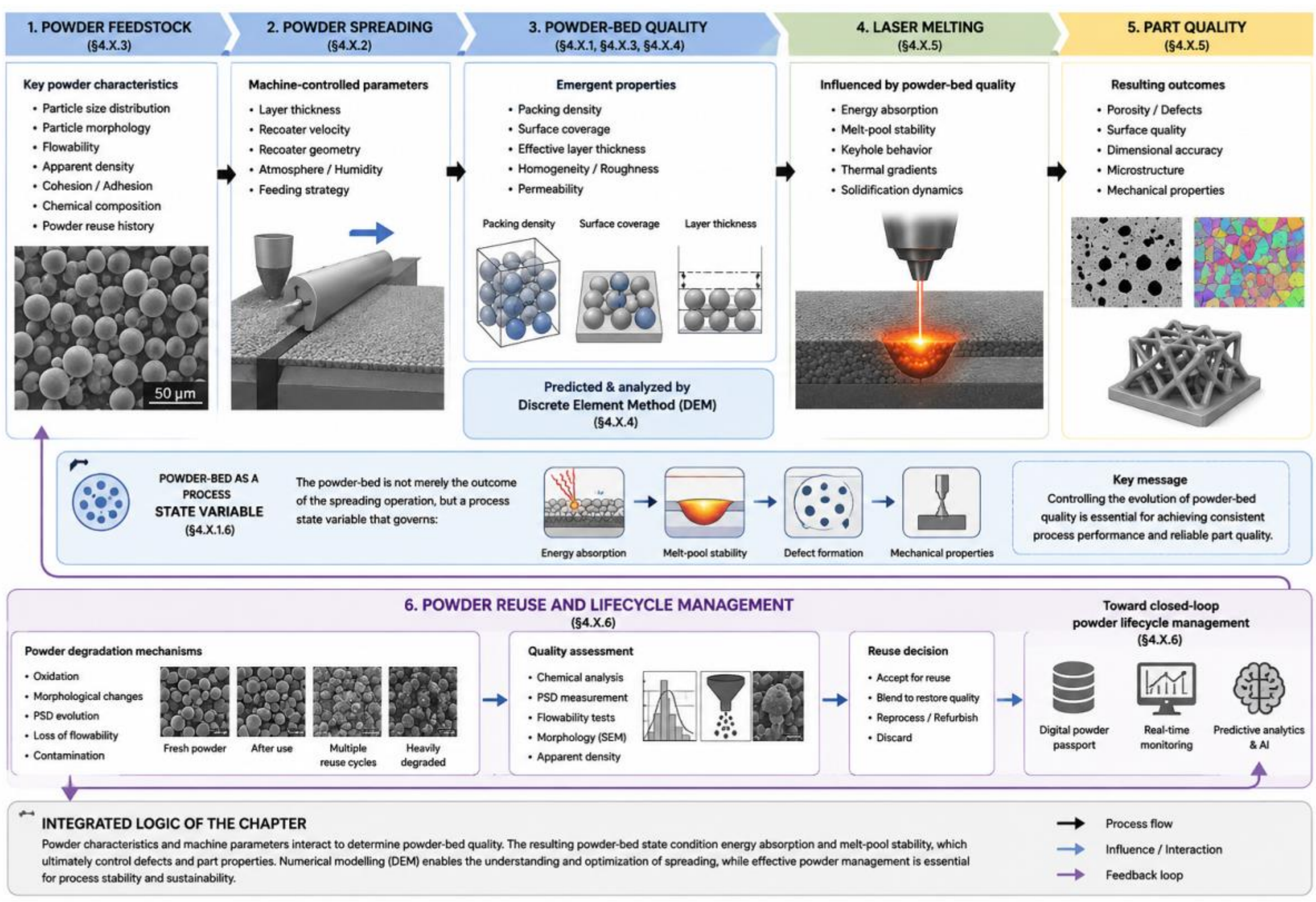

**Fig. 1:** Integrated logic of the chapter, linking powder feedstock characteristics, powder spreading, powder-bed quality, melting behaviour, part quality and powder lifecycle management.

The framework presented above highlights an important conceptual shift that has emerged in recent years within the additive manufacturing community. Traditionally, powder spreading was regarded as a preparatory operation whose role was limited to supplying material for the subsequent melting stage. However, increasing experimental and numerical evidence suggests that the powder-bed should be considered a process state variable rather than a simple process outcome. Powder-bed characteristics such as packing density, surface coverage, effective layer thickness, and spatial homogeneity collectively determine the thermal and optical conditions experienced during laser exposure, thereby influencing the effectiveness of energy transfer and the stability of the melt pool. This conceptual framework provides the foundation for the remainder of this chapter. The following sections will examine in greater detail the process parameters governing powder deposition, the influence of powder characteristics, the application of Discrete Element Method simulations to powder spreading, the relationship between powder-bed quality and melting behavior, and finally the industrial implications of powder handling and reuse strategies.

## 2.4 Process Parameters Governing Powder Deposition

The quality of a powder-bed is not solely determined by the characteristics of the feedstock powder but also by the process parameters governing the spreading operation. During Powder Bed Fusion (PBF), the deposition of a homogeneous powder layer results from a complex interaction between machine settings, powder properties, and environmental conditions. Even when high-quality powders are employed, inappropriate spreading parameters may generate heterogeneous powder-beds characterized by poor surface coverage, density fluctuations, particle segregation, and local thickness variations. The optimization of powder deposition therefore requires a comprehensive understanding of the mechanisms through which process parameters influence particle motion and powder-bed formation. Unlike laser parameters, whose influence is often directly observable through melt-pool behavior and part densification, spreading parameters act indirectly by modifying the initial state of the material before melting occurs. Nevertheless, their effect propagates throughout the entire manufacturing process and significantly contributes to process repeatability and final component quality. Among the numerous variables involved, layer thickness, recoater velocity, recoater design, environmental conditions, and powder feeding strategy are generally recognized as the most influential parameters governing powder deposition.

### *2.4.1 Layer Thickness*

Layer thickness is arguably the most important parameter controlling powder-bed formation. In industrial LPBF systems, nominal layer thicknesses typically range between 20 and 100 μm, depending on the material, machine architecture, and application requirements. From a geometrical perspective, thinner layers generally improve dimensional accuracy and reduce staircase effects on inclined surfaces. Since a smaller amount of material must be melted, thin layers often facilitate complete consolidation and contribute to improved surface finish. These advantages explain why aerospace and biomedical applications frequently employ layer thicknesses between 20 and 40 μm despite the associated increase in build time. However, reducing layer thickness introduces important challenges during spreading. As the gap between the recoater and the substrate approaches the size of the largest particles in the feedstock distribution, particle mobility becomes increasingly constrained. Large particles may become trapped beneath the recoater, leading to particle filtering, local disturbances, and jamming phenomena. Under these conditions, the deposited layer may no longer accurately reproduce the original particle size distribution of the feedstock. Numerical simulations have shown that when the ratio between nominal layer thickness and maximum particle diameter approaches unity, the probability of jamming increases significantly. Consequently, the quality of the powder-bed deteriorates, with reductions in both packing density and surface coverage. Conversely, thicker layers facilitate particle transport and generally improve spreading stability. The larger gap available beneath the recoater reduces the likelihood of particle blockage and enables a more efficient flow of material into the deposition zone. Nevertheless, increasing layer thickness also introduces disadvantages. More material must be melted during laser exposure, requiring higher energy input and potentially reducing dimensional accuracy. The selection of layer thickness therefore involves a trade-off between productivity, powder-bed quality, and process stability. Optimal values depend on both powder characteristics and the requirements of the final application.

### *2.4.2 Recoater Velocity*

The velocity of the recoating device exerts a profound influence on powder spreading dynamics. At relatively low spreading velocities, particle motion is primarily governed by gravity and frictional rearrangement mechanisms. Under these quasi-static conditions, particles have sufficient time to reorganize themselves while crossing the recoater gap, promoting improved packing density and more uniform deposition. As recoater velocity increases, inertial effects become progressively more important. Particles experience higher accelerations and may be transported over longer distances before settling onto the substrate. The resulting powder-bed tends to exhibit increased variability in both density and thickness. Experimental observations and DEM simulations consistently demonstrate that high spreading velocities can produce several undesirable effects:

- reduced packing density;
- lower surface coverage;
- increased layer roughness;
- stronger segregation phenomena;
- local powder accumulation zones;
- uncovered substrate regions.

The influence of recoater velocity becomes particularly critical when processing fine powders characterized by strong cohesive interactions. In such cases, rapid recoater motion may prevent complete particle rearrangement, leading to the formation of clusters and heterogeneous deposits. Despite these drawbacks, industrial systems often operate at relatively high spreading velocities to maximize productivity. Consequently, process optimization frequently requires balancing deposition quality against manufacturing throughput.

### *2.4.3 Recoater Geometry and Material*

The design of the recoating system directly affects powder transport and deposition mechanisms. Historically, rigid metallic blades represented the most common recoater configuration. Their widespread adoption was motivated by mechanical simplicity, dimensional precision, and robustness. However, rigid blades may generate significant mechanical interactions with protruding particles or previously consolidated features. These interactions can disturb the deposited layer and, in extreme cases, cause process interruptions. To overcome these limitations, many modern LPBF systems employ flexible polymer blades. Their deformability allows them to accommodate minor geometrical irregularities while reducing the risk of collisions with the build surface. Flexible recoaters therefore contribute to improved process robustness, particularly in large-format machines and long-duration builds. An alternative approach involves the use of rotating rollers. Unlike blades, rollers actively compact the powder during spreading, increasing local packing density. Experimental studies have shown that roller systems often produce denser and more homogeneous powder-beds than blade-based solutions. However, roller compaction also modifies particle rearrangement mechanisms and may increase segregation effects under certain conditions. Consequently, the optimal recoater design remains strongly dependent on material characteristics and process objectives. Recent developments have introduced hybrid systems combining multiple spreading mechanisms or integrating vibration-assisted deposition strategies. These solutions aim to further improve powder-bed homogeneity and reduce sensitivity to powder characteristics.

### *2.4.4 Environmental Conditions*

Although powder spreading is often viewed as a purely mechanical process, environmental conditions significantly influence powder behavior. Humidity represents one of the most critical variables. Moisture adsorption promotes the formation of capillary bridges between neighboring particles, increasing cohesion and reducing flowability. This phenomenon is particularly pronounced for fine metallic powders, where cohesive forces are already comparable to particle weight. The resulting increase in cohesion may lead to:

- particle agglomeration;
- reduced powder mobility;
- poor surface coverage;

- density fluctuations;
- spreading instabilities.

Consequently, industrial LPBF facilities typically maintain controlled humidity environments during powder storage and handling operations. Oxygen concentration also influences powder quality. Although inert gas atmospheres are primarily employed to prevent oxidation during melting, oxygen exposure may progressively modify powder characteristics during repeated reuse cycles. Changes in surface chemistry affect both spreading behavior and subsequent melting performance. Gas flow conditions within the build chamber represent another relevant factor. Modern LPBF systems employ carefully designed gas circulation systems to remove spatter and process emissions generated during laser exposure. However, these flows may also interact with loose powder particles, influencing their redistribution and contributing to local variations in powder-bed morphology. Finally, chamber temperature may affect powder spreading by modifying particle friction coefficients and material properties. While these effects are generally less pronounced than those associated with humidity and cohesion, they become increasingly relevant in high-temperature PBF systems.

#### *2.4.5 Powder Feeding Strategy and Reservoir Conditions*

The powder feeding system constitutes an often-overlooked aspect of powder deposition. Before reaching the build platform, particles are stored within a feed reservoir where they may experience segregation and rearrangement. The state of the powder within this reservoir directly influences subsequent spreading behavior. DEM investigations have shown that the movement of the recoater continuously modifies the composition of the powder heap accumulated in front of the blade. Larger particles tend to remain concentrated near the front of the heap, whereas smaller particles preferentially migrate into void spaces and are deposited earlier during spreading. As a result, the characteristics of the deposited layer may differ from those of the original feedstock. This effect becomes more pronounced when large numbers of layers are deposited or when powder reuse strategies are implemented. The amount of powder available in the feed reservoir also affects spreading consistency. Insufficient powder supply may generate short-feed conditions characterized by incomplete layer coverage. Excessive powder feed, on the other hand, increases material handling requirements and may promote segregation phenomena. The optimization of powder feeding strategies therefore represents an important component of process control, particularly in industrial production environments.

#### *2.4.6 Interactions Between Process Parameters*

One of the main challenges in powder spreading optimization is that process parameters rarely act independently. For example, the influence of recoater velocity depends strongly on layer thickness. High recoater velocitys may be acceptable when thick layers are employed but become problematic when thin layers are combined with cohesive powders. Similarly, the optimal layer thickness depends on particle size distribution. A value that performs well for one feedstock may lead to severe jamming phenomena when applied to a different powder. Environmental conditions further complicate this picture by modifying particle cohesion and flowability. As a result, parameter optimization should always be approached as a multidimensional problem rather than a sequence of independent adjustments. This complexity explains the increasing adoption of integrated numerical-experimental methodologies in additive manufacturing process development.

# 3 Toward Optimized Powder Deposition

The ultimate objective of powder spreading optimization is the generation of powder-beds characterized by:

- high packing density;
- complete surface coverage;
- uniform layer thickness;
- limited segregation;
- high repeatability.

Achieving these conditions requires balancing competing objectives. Low spreading velocities and moderate layer thicknesses generally maximize powder-bed quality but reduce productivity. Conversely, aggressive operating conditions increase manufacturing throughput while potentially compromising process stability. The growing availability of DEM simulations, in situ monitoring systems, and data-driven optimization techniques is progressively enabling a more rational selection of spreading parameters. These tools allow engineers to identify optimal operating windows capable of simultaneously satisfying productivity and quality requirements. As will be discussed in the following sections, the influence of process parameters cannot be fully understood without considering the characteristics of the powder itself. The interaction between process conditions and feedstock properties ultimately determines the quality of the deposited layer and the success of the entire Powder Bed Fusion process.

## 3.1 Influence of Powder Characteristics on Layer Deposition

The quality of the powder-bed generated during Powder Bed Fusion (PBF) processes is strongly dependent on the intrinsic characteristics of the feedstock powder. While process parameters such as layer thickness and recoater velocity govern the dynamics of powder spreading, the response of the granular material is ultimately determined by particle-scale properties. Consequently, powder characteristics directly influence flowability, packing density, surface coverage, absorptivity, and process repeatability. The growing industrial adoption of LPBF technologies has led to increasing attention toward powder engineering and powder qualification. Modern additive manufacturing powders are no longer regarded simply as raw materials but rather as engineered feedstocks specifically designed to satisfy stringent requirements regarding morphology, particle size distribution, chemical composition, and flow behavior. Among the various factors influencing powder deposition, particle size distribution, particle morphology, density, cohesion, surface chemistry, and flowability are generally recognized as the most important. These characteristics collectively determine the ability of the powder to form homogeneous layers and therefore establish the initial conditions for the subsequent melting process.

### *3.1.1 Particle Size Distribution*

Particle size distribution (PSD) is widely considered the most influential powder characteristic in Powder Bed Fusion technologies. Commercial LPBF powders are typically characterized by particle diameters ranging from approximately 15 to 63 μm, although broader or narrower distributions may be adopted depending on the alloy system and application requirements. The PSD affects nearly every stage of the manufacturing process, including powder handling, spreading, energy absorption, and melting. A broad particle size distribution generally promotes improved packing density because smaller particles can occupy the interstitial spaces between larger particles. This effect reduces the overall void fraction of the deposited layer and increases the amount of material available for melting within a given volume. From a geometrical perspective, the presence of fine particles allows a more efficient filling of local voids generated during particle rearrangement. Consequently, broad PSDs often produce denser powder-beds than narrowly distributed powders. However, excessively broad distributions may also generate undesirable effects. Significant differences in particle size increase the likelihood of segregation during handling and spreading operations. Larger particles tend to migrate differently from smaller ones, leading to local variations in composition, density, and absorptivity throughout the powder-bed. DEM investigations have clearly demonstrated that PSD influences both packing density and surface coverage. Numerical simulations indicate that the relationship between layer thickness and maximum particle diameter plays a critical role in determining spreading quality. When the nominal layer thickness becomes comparable to the size of the largest particles, particle filtering and jamming phenomena may occur, resulting in reduced powder-bed homogeneity. These observations highlight the need for careful matching between PSD and process parameters.

### *3.1.2 Particle Morphology and Sphericity*

Particle morphology represents another fundamental characteristic influencing powder-bed formation. Most metallic powders employed in additive manufacturing are produced through gas atomization processes. This manufacturing route generates particles with a high degree of sphericity, which is generally considered desirable for powder spreading applications. Spherical particles exhibit several advantages:

- reduced interparticle friction;
- improved flowability;
- enhanced particle rearrangement;
- more homogeneous powder deposition;
- lower tendency toward mechanical interlocking.

As a result, powders characterized by high sphericity typically produce denser and more uniform powder-beds. Conversely, irregular particles increase frictional interactions and hinder particle mobility. The resulting reduction in flowability may generate heterogeneous layers characterized by local density fluctuations and poor surface coverage. In addition to particle shape itself, the presence of satellites also deserves particular attention. Satellites are small particles attached to larger particles during atomization or subsequent processing. Although their influence on particle size distribution may be limited, satellites significantly increase surface roughness and modify contact mechanics between neighboring particles. The accumulation of satellites is particularly relevant during powder reuse because repeated thermal exposure and handling operations may progressively increase their concentration within the feedstock.

### *3.1.3 Surface Texture and Contact Mechanics*

The behavior of additive manufacturing powders is governed by interactions occurring at particle contact points. Even when particles appear nearly spherical at the macroscopic level, microscopic surface features strongly influence their mechanical behavior. Surface roughness modifies frictional forces, contact stiffness, and cohesive interactions between particles. Rough particle surfaces generally increase resistance to relative motion and reduce flowability. This effect becomes particularly important for fine powders because surface-related forces scale differently than gravitational forces. As particle size decreases, the ratio between surface area and volume increases substantially. Consequently, the influence of contact mechanics becomes progressively more important relative to particle weight. This observation explains why powders characterized by similar particle size distributions may exhibit significantly different spreading behavior depending on their surface characteristics.

### *3.1.4 Cohesion and Interparticle Forces*

One of the defining features of additive manufacturing powders is the increasing importance of cohesive interactions. For coarse granular materials, gravity typically dominates particle behavior. However,

LPBF powders frequently exhibit particle diameters below 100 μm, a range in which attractive forces become comparable to particle weight. Several mechanisms contribute to cohesion:

- van der Waals interactions;
- electrostatic forces;
- capillary bridges;
- chemical surface interactions.

These forces promote particle clustering and agglomeration, altering powder flow behavior and reducing spreadability. Experimental studies consistently demonstrate that cohesion increases as particle size decreases. Consequently, powders containing large fractions of fine particles often exhibit poorer flowability despite their potential advantages regarding packing density. The importance of cohesion has also been highlighted by DEM investigations. Calibration procedures frequently identify cohesive energy density as one of the most influential parameters controlling powder spreading behavior. Simulations neglecting cohesive interactions generally overestimate powder mobility and fail to reproduce experimentally observed powder-bed characteristics. The balance between packing efficiency and cohesion therefore represents one of the central challenges in powder design for additive manufacturing.

#### *3.1.5 Flowability*

Flowability is commonly regarded as the most comprehensive descriptor of powder performance during spreading. In practice, flowability reflects the combined influence of particle size distribution, morphology, density, surface texture, and cohesion. Powders exhibiting good flowability generally spread more uniformly and generate powder-beds characterized by improved homogeneity and repeatability. Several experimental methods have been developed to quantify flowability, including:

- Hall flow tests;
- Carney flow tests;
- angle-of-repose measurements;
- avalanche testing;
- rotating drum methods;
- shear-cell characterization.

Although these techniques provide valuable information, none of them fully reproduces the dynamic conditions encountered during LPBF spreading. This limitation arises because powder deposition occurs under highly constrained geometries involving layer thicknesses often comparable to particle diameters. Consequently, conventional flowability metrics should be interpreted as indicators rather than direct predictors of spreading performance. For this reason, process-specific testing procedures and DEM simulations are increasingly employed alongside traditional powder characterization methods.

### 3.2 Packing Density and Powder-Bed Formation

Packing density represents the most important macroscopic consequence of powder characteristics. A powder-bed characterized by high packing density contains a larger quantity of material per unit volume and exhibits reduced void content. Such conditions generally promote improved thermal continuity and more efficient energy absorption during melting. Experimental and numerical studies consistently demonstrate that packing density depends strongly on PSD, morphology, and cohesion. Broad PSDs typically increase packing efficiency by facilitating interstitial filling. Spherical particles enhance particle rearrangement and improve compaction. Conversely, excessive cohesion limits particle mobility and reduces the achievable solid volume fraction. The optimization of powder characteristics therefore frequently aims to maximize packing density while maintaining acceptable flowability. This objective is particularly important because packing density establishes a direct connection between powder engineering and process performance.

### 3.3 Powder Characteristics and Energy Absorption

The influence of powder characteristics extends well beyond the spreading stage. The optical and thermal behavior of a powder-bed depends strongly on the arrangement of particles within the deposited layer. Particle size distribution, packing density, and surface morphology collectively determine how incident laser radiation interacts with the material. Dense powder-beds promote multiple reflections of incoming radiation between neighboring particles. This mechanism increases the probability of energy absorption and reduces reflection losses. The effect is especially important for highly reflective alloys such as aluminum-based systems. Several investigations have shown that improvements in powder-bed density can substantially enhance absorptivity and reduce the energy required for complete melting. Consequently, powder characteristics indirectly influence:

- melt-pool stability;
- porosity formation;
- process efficiency;
- final component properties.

This relationship establishes a direct link between feedstock engineering and manufacturing performance.

### 3.4 Influence of Powder Reuse on Powder Characteristics

The progressive industrialization of LPBF technologies has increased interest in powder recycling strategies.

- Repeated reuse cycles inevitably modify powder characteristics through:
- sieving operations;
- thermal exposure;
- oxidation;
- particle agglomeration;
- satellite formation.

Studies performed on AlSi10Mg powders have shown that PSD gradually shifts toward smaller particle diameters due to the removal of coarse particles and agglomerates during sieving. At the same time, repeated processing may alter particle morphology and surface chemistry. Although these modifications are often relatively small, they may influence flowability, apparent density, and powder-bed formation. Interestingly, several investigations demonstrate that properly managed powders can be reused multiple times without significant degradation of mechanical properties. Nevertheless, maintaining stable powder characteristics remains essential for ensuring process repeatability and quality consistency. For this reason, modern industrial facilities increasingly implement systematic powder monitoring protocols including PSD measurements, density characterization, chemical analysis, and flowability assessment.

### 3.5 Powder Engineering as a Process Design Tool

Traditionally, powder characterization was viewed as a material qualification activity performed prior to manufacturing. Modern research increasingly suggests a different perspective. Because powder characteristics directly influence powder-bed formation, energy absorption, and defect formation, they should be regarded as process variables rather than merely material descriptors. This shift in perspective has given rise to the concept of powder engineering for additive manufacturing. Instead of adapting process parameters to available powders, powder design itself becomes part of process optimization. The integration of advanced atomization technologies, powder characterization methods, DEM simulations, and machine-learning tools is progressively enabling the development of feedstocks specifically tailored for additive manufacturing applications. Ultimately, the ability to engineer powder characteristics represents one of the most promising pathways toward improved process robustness, reduced variability, and enhanced component performance in future Powder Bed Fusion systems.

# 4 DEM-Based Modelling of Powder Spreading

## 4.1 Discrete Element Method (DEM) for Powder Spreading Simulations

The Discrete Element Method (DEM) has become one of the most widely adopted numerical approaches for investigating powder spreading phenomena in Powder Bed Fusion (PBF) processes. Unlike continuum-based methods, which treat granular materials as homogeneous media, DEM explicitly models the motion and interaction of individual particles, making it particularly suitable for additive manufacturing powders whose dimensions are comparable to the deposited layer thickness. In DEM simulations, each particle is represented as a discrete entity whose translational and rotational motion is governed by Newton’s equations of motion. At every computational time step, contact forces arising from particle–particle and particle–wall interactions are calculated and used to update particle positions and velocities. This approach enables the direct observation of particle rearrangement, segregation, jamming, clustering, and powder-bed formation mechanisms that are difficult to investigate experimentally. The contact interactions are typically described using the Hertz–Mindlin contact model, which accounts for elastic deformation, frictional effects, and energy dissipation during collisions. Since additive manufacturing powders generally exhibit particle diameters below 100 µm, cohesive forces become increasingly relevant and must often be incorporated into the model. Consequently, DEM simulations commonly include additional force contributions associated with van der Waals attraction, rolling resistance, and simplified cohesion energy density formulations. The inclusion of these effects is essential for accurately reproducing the flow behaviour of fine metallic powders. A critical aspect of DEM modelling is parameter calibration. Material properties such as density and elastic modulus can generally be obtained from literature or experimental measurements, whereas friction coefficients, rolling resistance coefficients, and cohesive parameters are typically calibrated against experimentally measurable quantities. Common calibration targets include the angle of repose, angle of slip, apparent density, tapped density, and powder heap geometry. Multi-objective calibration strategies are often preferred because they provide more robust parameter sets capable of reproducing different aspects of powder behaviour. The numerical domain used for powder spreading simulations generally reproduces the key components of a PBF machine, including the powder reservoir, build platform, and recoating device. Powder particles are initially generated within a feed region and allowed to settle under gravity. The recoater motion is then simulated by imposing a prescribed displacement to a rigid blade, flexible blade, or roller system. During spreading, particle trajectories, contact forces, and local packing conditions are continuously monitored. The resulting powder-bed is typically evaluated using quantitative metrics such as packing density, surface coverage, effective layer thickness, surface roughness, and particle segregation indices. These parameters provide valuable information regarding the quality and homogeneity of the deposited layer and establish a direct connection between powder characteristics, spreading conditions, and subsequent melting behaviour. Although DEM simulations remain computationally demanding, they have become an essential tool for understanding powder-bed formation and optimizing process parameters in metal additive manufacturing. Their capability to capture particle-scale phenomena makes DEM particularly valuable for investigating the influence of powder characteristics, recoating strategies, and environmental conditions on powder-bed quality, thereby supporting the development of more robust and reliable Powder Bed Fusion processes.

A representative DEM workflow employed for powder-spreading investigations is shown in Fig. 2.

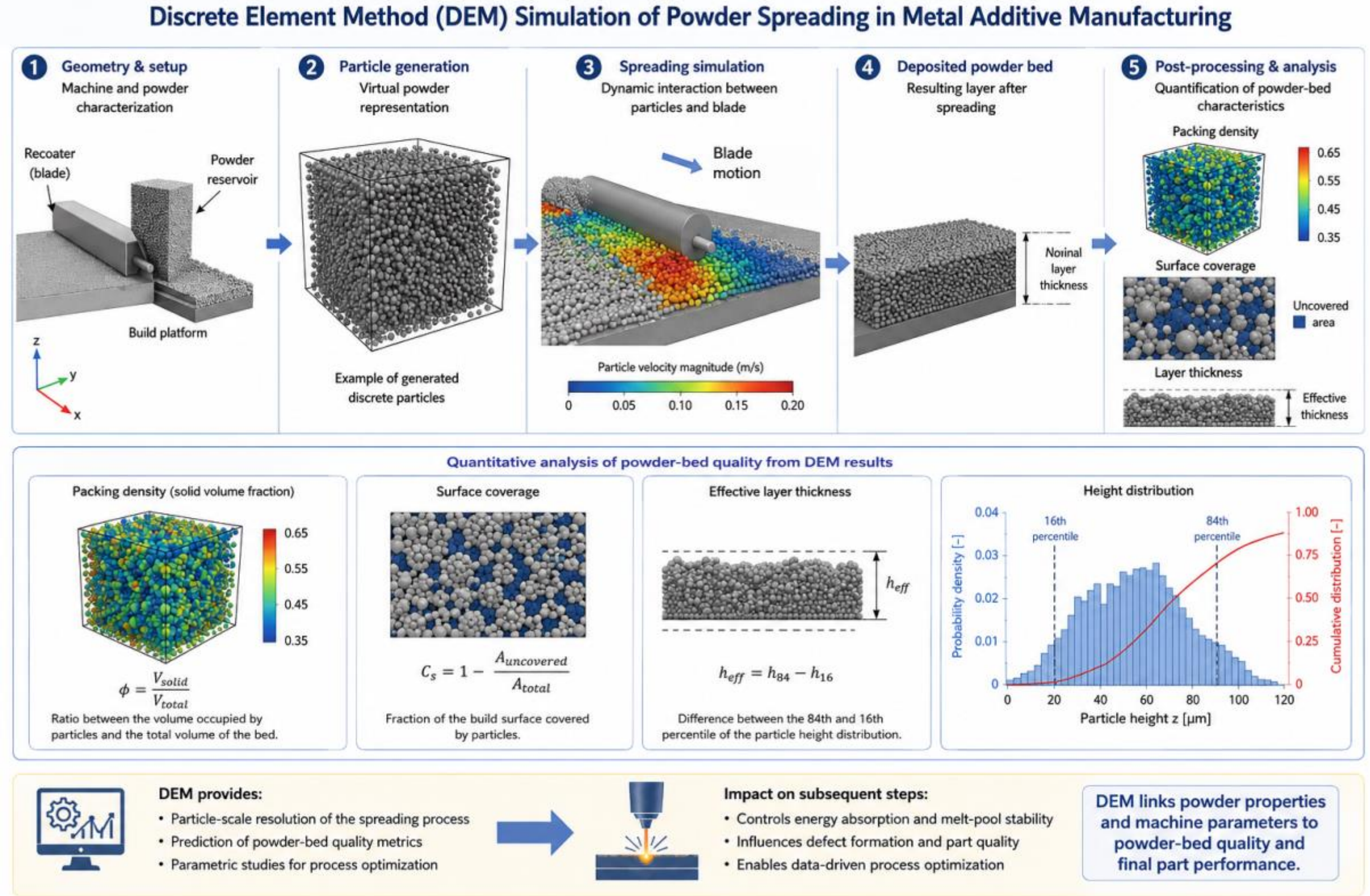

**Fig. 2:** DEM-based workflow for powder spreading simulations, including geometry definition, particle generation, spreading simulation and quantitative powder-bed quality metrics.

## 5 Powder-Bed Quality as an Emergent Property

The concept of powder-bed quality extends beyond the evaluation of individual powder characteristics. Modern studies increasingly indicate that powder-bed quality should be regarded as an emergent property resulting from the interaction between powder characteristics, machine parameters, and environmental conditions. Particle size distribution, morphology, cohesion, recoater velocity, layer thickness, and humidity do not act independently but collectively determine the final arrangement of particles within the deposited layer. Consequently, powder-bed quality emerges as the outcome of a complex granular system and motivates the adoption of particle-scale modelling approaches such as the Discrete Element Method. The increasing recognition of powder spreading as a critical stage in Powder Bed Fusion (PBF) technologies has stimulated the development of advanced numerical tools capable of describing powder-bed formation at the particle scale. Among the available modelling approaches, the Discrete Element Method (DEM) has emerged as the most widely adopted framework for investigating powder spreading dynamics, owing to its ability to explicitly represent individual particles and their interactions. Unlike continuum approaches, which treat the powder-bed as a homogeneous medium, DEM resolves the motion of every particle within the computational domain, providing direct access to local phenomena such as particle rearrangement, segregation, jamming, cohesion, and powder-bed densification. The relevance of DEM in additive manufacturing stems from the intrinsic nature of the spreading process. Powder deposition is fundamentally a granular flow problem involving particles whose dimensions are comparable to the layer thickness itself. Under these conditions, continuum assumptions become questionable, particularly when local particle-scale interactions govern the final arrangement of the deposited layer. DEM overcomes these limitations by solving the equations of motion of each particle individually, thereby capturing the stochastic nature of powder-bed formation.

## 5.1 Powder Bed Quality as a Process Variable

The relationship between powder characteristics, powder-bed quality, melting behavior and final component performance is summarized in Fig. 3.

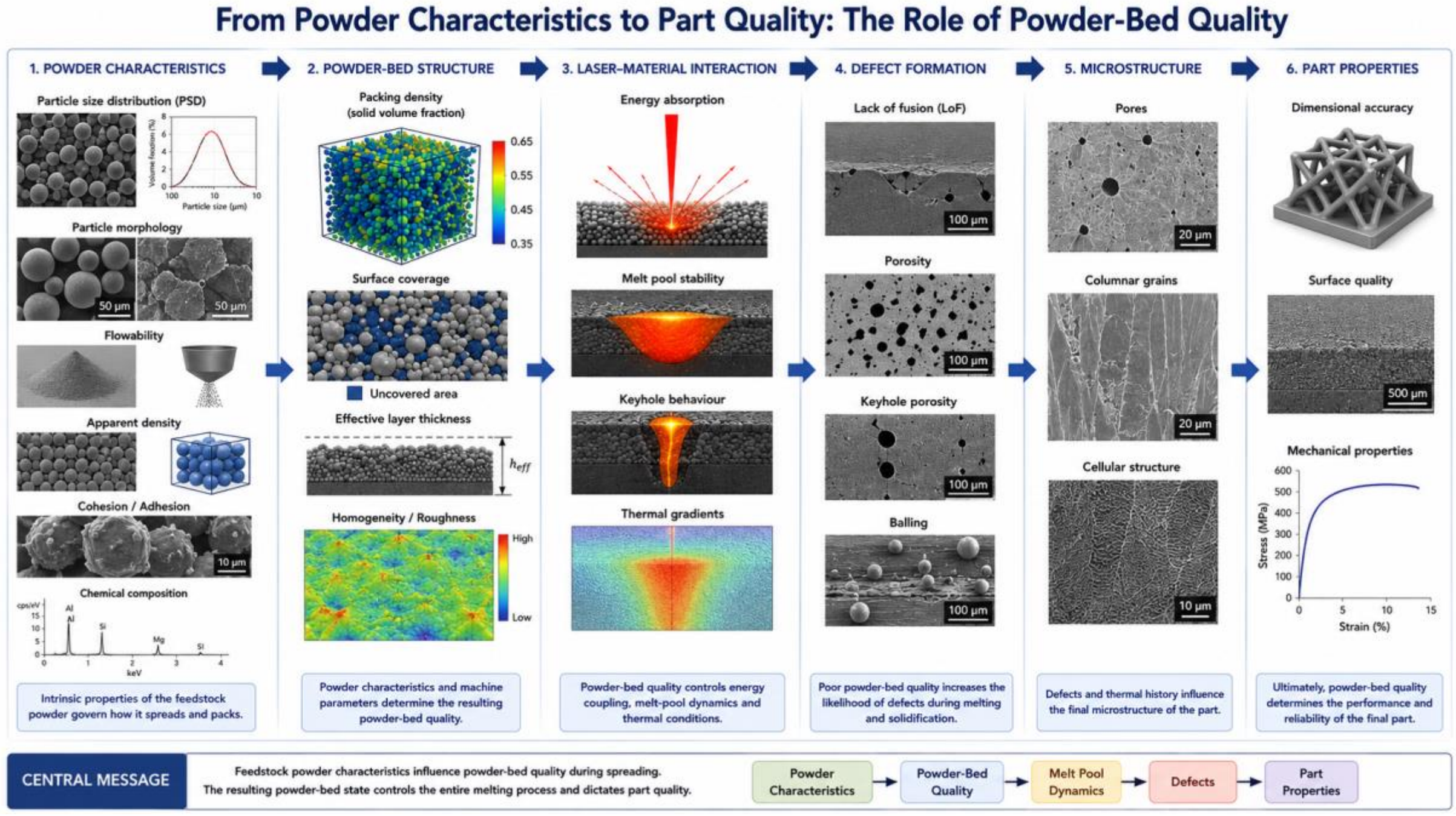

**Fig. 3:** Relationship between powder characteristics, powder-bed structure, laser-material interaction, defect formation, microstructure development and final component properties.

Traditionally, the powder-bed has been treated as a passive substrate upon which the laser or electron beam acts. However, this perspective overlooks the fact that powder-bed characteristics such as packing density, layer thickness uniformity, particle size distribution, and surface coverage directly influence the absorption, distribution, and dissipation of thermal energy. Unlike bulk metallic materials, powder-beds are highly heterogeneous systems composed of solid particles separated by void spaces filled with inert gas. Consequently, their thermal and optical properties differ significantly from those of the corresponding fully dense material. The amount of energy effectively absorbed by the powder-bed depends not only on the intrinsic absorptivity of the material but also on the geometrical arrangement of particles within the deposited layer.

Variations in powder-bed quality can therefore generate substantial differences in local thermal conditions even when identical laser parameters are employed. This observation explains why components produced using nominally identical process settings may exhibit significant variability in density, surface quality, and mechanical properties.

## 5.2 Influence of Packing Density on Energy Absorption

Among all powder-bed characteristics, packing density is generally recognized as the most influential parameter governing energy transfer during melting. A dense powder-bed contains a larger quantity of material per unit volume and exhibits reduced void content. From an optical perspective, this configuration promotes multiple reflections of the incident radiation between neighboring particles. Instead of being reflected away from the process zone, a larger fraction of the incoming energy remains trapped within the powder layer and is progressively absorbed by the material. The importance of this mechanism becomes particularly evident in highly reflective alloys such as aluminum-based systems. Aluminum alloys exhibit relatively low absorptivity at the laser wavelengths commonly employed in LPBF and are therefore characterized by higher energy requirements compared to many steels, titanium alloys, or nickel-based superalloys. Several experimental and numerical studies have demonstrated that improvements in powder-bed packing density can significantly enhance effective energy absorption. DEM-assisted investigations have shown that optimization of spreading conditions may enable the production of dense components using substantially lower nominal energy densities than those traditionally reported in the literature. These findings suggest that powder-bed optimization can be considered an indirect strategy for reducing process energy consumption.

### 5.3 Thermal Conductivity and Heat Transfer in Powder Beds

The thermal behavior of powder-beds differs substantially from that of fully dense materials. Because heat must be transferred through a network of discrete particle contacts, the effective thermal conductivity of a powder layer is considerably lower than the conductivity of the bulk alloy. The number and quality of particle-particle contacts strongly depend on packing density. Dense powder-beds contain a larger number of contact points, thereby facilitating heat conduction through the layer. Conversely, loosely packed beds exhibit poor thermal connectivity and higher thermal resistance. This effect influences both the heating and cooling stages of the process. During laser exposure, local temperature gradients become more pronounced in poorly packed regions. Following melting, variations in thermal conductivity affect solidification rates and microstructural evolution. Consequently, spatial heterogeneities in powder-bed density may generate localized variations in grain morphology, residual stress distribution, and mechanical performance.

### 5.4 Powder Bed Quality and Melt Pool Stability

The melt pool represents the central feature of LPBF processing. Its geometry, stability, and evolution determine the quality of the consolidated material. Powder-bed heterogeneities introduce fluctuations in energy absorption that directly affect melt-pool dynamics. Regions characterized by low packing density may receive insufficient energy to achieve complete melting, while excessively dense regions may alter local heat accumulation and fluid-flow conditions. A stable melt pool requires relatively uniform thermal conditions along the scanning trajectory. Powder-bed defects such as uncovered areas, density gradients, particle clusters, or local thickness variations disrupt this stability and increase the probability of process-induced defects. The relationship between powder-bed quality and melt-pool behavior is particularly relevant when operating near the boundaries of the process window. Under such conditions, even relatively small variations in absorptivity or thermal conductivity may determine whether complete fusion is achieved.

### 5.5 Defects

**Lack-of-fusion (LOF) defects**: represent one of the most common and detrimental defect categories in Powder Bed Fusion processes. These defects originate when insufficient energy is delivered to fully melt the powder particles and remelt the underlying layer. While inadequate laser parameters are often considered the primary cause of LOF formation, powder-bed quality frequently plays an equally important role. Poor surface coverage, excessive layer thickness, particle agglomeration, and low packing density reduce local energy absorption and hinder complete consolidation. As a result, irregularly shaped voids remain trapped between neighboring melt tracks or successive layers. LOF defects are typically characterized by elongated morphologies and sharp corners, making them particularly harmful from a mechanical perspective. Their geometry promotes stress concentration and significantly reduces fatigue resistance. Experimental observations consistently show that LOF defects become more prevalent when spreading conditions deteriorate or when powder characteristics deviate from their optimal range.

**Keyhole Porosity and Excessive Energy Input**: While insufficient energy leads to lack-of-fusion, excessive local energy input may generate a different class of defects known as keyhole porosity. Keyholing occurs when intense vaporization generates a deep and unstable cavity within the molten material. The subsequent collapse of this cavity traps gas bubbles that become spherical pores after solidification. Although keyhole defects are primarily associated with laser parameters, powder-bed quality can influence their formation indirectly. Variations in absorptivity caused by local differences in packing density may alter the effective energy delivered to the material, thereby shifting the process toward unstable operating conditions. Consequently, identical laser settings may produce different outcomes depending on the quality of the deposited powder layer.

**Balling and Melt Track Instabilities**: Another defect category frequently associated with powder-bed quality is balling. Balling phenomena occur when the molten material loses stability and breaks into discrete droplets rather than forming continuous melt tracks. This behavior is governed by a complex interaction between surface tension, viscosity, wetting behavior, and thermal gradients. Powder-bed heterogeneities contribute to balling by generating local fluctuations in melt-pool dimensions and temperature. Surface irregularities resulting from poor spreading may further amplify these instabilities, leading to rough surfaces and incomplete consolidation. Balling is particularly problematic because it affects not only the current layer but also all subsequent layers deposited on top of the defective region.

**Surface Roughness and Dimensional Accuracy:** The influence of powder-bed quality extends beyond internal porosity and directly affects the external characteristics of the manufactured component. Non-uniform powder deposition generates local variations in layer thickness and melt-pool dimensions, leading to increased surface roughness and dimensional deviations. In extreme cases, powder-bed defects may accumulate layer after layer and produce macroscopic geometric distortions. Because additive manufacturing relies on the repeated deposition of hundreds or thousands of layers, even small spreading imperfections can progressively affect component accuracy. Surface roughness is particularly sensitive to powder characteristics such as particle size distribution and morphology. Coarse particles, satellites, and partially melted agglomerates frequently contribute to rough surface formation and reduced dimensional fidelity.

## 5.6 Influence on Mechanical Properties

The framework presented above highlights an important conceptual shift that has emerged in recent years within the additive manufacturing community. Traditionally, powder spreading was regarded as a preparatory operation whose role was limited to supplying material for the subsequent melting stage. However, increasing experimental and numerical evidence suggests that the powder-bed should be considered a process state variable rather than a simple process outcome. Powder-bed characteristics such as packing density, surface coverage, effective layer thickness, and spatial homogeneity collectively determine the thermal and optical conditions experienced during laser exposure, thereby influencing the effectiveness of energy transfer and the stability of the melt pool. The mechanical properties of LPBF components are ultimately controlled by the defects and microstructural features generated during processing. Powder-bed quality influences these properties through multiple pathways. Improved packing density promotes enhanced densification, reduced porosity, and more stable melt-pool dynamics. These factors generally translate into increased tensile strength, improved ductility, higher fatigue resistance, and reduced scatter in mechanical performance. Conversely, powder-bed heterogeneities increase the likelihood of defect formation and may generate localized variations in microstructure. Such variations contribute to anisotropy, property scatter, and reduced reliability. The strong correlation between powder-bed quality and mechanical performance further emphasizes the importance of considering spreading as a primary process variable rather than a secondary preparatory operation.

## 6 Towards Integrated Process Optimization: concluding remarks and future research directions

The growing availability of DEM simulations, in situ monitoring systems, and advanced characterization techniques is enabling a more integrated approach to process optimization. Instead of optimizing laser parameters independently, modern strategies increasingly consider the entire manufacturing chain, linking powder characteristics, spreading conditions, powder-bed quality, melt-pool behavior, defect formation, and final component performance. Within this framework, powder-bed quality emerges as the key intermediary variable connecting feedstock properties and process parameters to the final product. The optimization of powder spreading therefore represents one of the most promising pathways toward improved process robustness, reduced energy consumption, and enhanced component reliability in future Powder Bed Fusion technologies. Powder spreading and layer deposition constitute fundamental stages of Powder Bed Fusion (PBF) technologies and play a significantly more important role than traditionally assumed in additive manufacturing process design. Although research efforts have historically focused on laser–material interaction, melt pool dynamics, and solidification phenomena, growing experimental and numerical evidence demonstrates that the quality of the deposited powder layer largely determines the effectiveness of all subsequent process stages. The powder-bed acts as the physical medium through which energy is transferred to the material, establishing a direct connection between feedstock characteristics, process parameters, and final component performance. The present chapter has highlighted how powder-bed quality emerges from the complex interaction between powder properties and spreading conditions. Process parameters such as layer thickness, recoater velocity, and recoater geometry govern the kinematics of particle rearrangement, while powder characteristics including particle size distribution, morphology, cohesion, and flowability determine the ability of the feedstock to form homogeneous layers. The resulting powder-bed is characterized by properties such as packing density, surface coverage, effective layer thickness, and spatial uniformity, all of which directly influence melting behavior and defect formation. Particular attention has been devoted to the role of particle-scale phenomena in determining powder-bed quality. Experimental observations and Discrete Element Method (DEM) simulations have shown that segregation, jamming, clustering, and density fluctuations naturally emerge during spreading and may significantly alter local process conditions. These phenomena are especially relevant in modern LPBF systems, where layer thicknesses are often comparable to particle dimensions and where even minor variations in powder-bed quality can influence energy absorption and melt-pool stability. One of the most important conclusions arising from recent research is that powder-bed quality should be considered a process variable in its own right. Packing density, for example, has been shown to influence absorptivity, thermal conductivity, and energy-transfer efficiency. Dense powder-beds promote multiple internal reflections of incident radiation, increasing energy absorption and facilitating stable melting conditions. Consequently, optimization of the spreading process may contribute not only to improved component quality but also to reduced energy consumption and enhanced process sustainability. The chapter has also emphasized the growing importance of powder lifecycle management. Industrial adoption of additive manufacturing relies heavily on powder recycling strategies aimed at reducing production costs and minimizing material waste. Although numerous studies demonstrate that powders can often be reused multiple times without significant degradation of component properties, physical and chemical evolution of the feedstock inevitably occurs. Changes in particle size distribution, oxidation state, morphology, and flowability progressively modify powder-bed formation and must therefore be monitored through appropriate qualification procedures. The transition from empirical powder reuse practices to scientifically controlled powder management systems represents a critical step toward robust industrial implementation of PBF technologies. Despite the considerable progress achieved over the last decade, several scientific and technological challenges remain open. A first challenge concerns the development of more realistic numerical models. Current DEM simulations have greatly improved the understanding of powder spreading mechanisms; however, computational limitations still require significant simplifications regarding particle shape, domain size, and process duration. Future developments should focus on multi-layer simulations, realistic particle morphologies, and stronger integration with thermal and fluid-dynamic models. Coupled DEM–CFD and DEM–thermal frameworks are expected to provide a more comprehensive description of the relationships between powder-bed formation, melt-pool evolution, and final component quality. A second challenge relates to in situ monitoring and process control. Although significant advances have been made in melt-pool monitoring, direct observation of powder spreading remains relatively limited in industrial systems. Future machine architectures will likely integrate advanced sensing technologies capable of measuring powder-bed density, surface coverage, and layer uniformity in real time. Such capabilities would enable immediate detection of spreading anomalies and facilitate adaptive process control strategies. Artificial intelligence and machine learning represent another promising research direction. The large datasets generated by numerical simulations, process monitoring systems, and powder characterization techniques create opportunities for the development of predictive models capable of correlating powder properties, spreading conditions, and component performance. Machine-learning-assisted optimization may substantially reduce experimental effort and accelerate process qualification for new materials. The concept of the digital twin is also expected to play a central role in the future of additive manufacturing. By combining DEM simulations, process monitoring data, powder characterization results, and thermal models within a unified computational framework, digital twins could enable real-time prediction of powder-bed quality and process outcomes. Such systems would represent a major step toward autonomous manufacturing environments characterized by enhanced reliability and reduced operator dependency. Finally, sustainability considerations are likely to become increasingly important. Powder production is an energy-intensive process, and feedstock materials often represent a significant fraction of the environmental footprint associated with additive manufacturing. Future research should therefore focus on improving powder reuse strategies, reducing material waste, and developing standardized methodologies for powder qualification and lifecycle assessment. The integration of sustainability metrics into process optimization frameworks may contribute to the broader adoption of additive manufacturing in environmentally conscious production systems. In conclusion, powder spreading and layer deposition should no longer be regarded as secondary operations preceding melting. Instead, they constitute critical process stages that strongly influence energy absorption, defect formation, process repeatability, and final component performance. Continued advances in powder engineering, numerical modelling, process monitoring, and digital manufacturing technologies will further enhance understanding and control of powder-bed formation, ultimately contributing to the maturation and industrialization of Powder Bed Fusion as a robust and sustainable manufacturing technology. Future advances in Powder Bed Fusion will depend not only on the optimization of laser-material interaction but increasingly on the capability to understand, monitor, and control powder-bed formation throughout the entire powder lifecycle. The integration of powder engineering, advanced process monitoring, numerical modelling, and intelligent powder management systems will play a pivotal role in transforming Powder Bed Fusion into a fully mature, reliable, and sustainable manufacturing technology.

## References

[1] Cundall, P.A., Strack, O.D.L. (1979). A discrete numerical model for granular assemblies. *Géotechnique*, 29, 47–65. https://doi.org/10.1680/geot.1979.29.1.47

[2] Lampitella, V., Trofa, M., Astarita, A., D'Avino, G. (2021). Discrete Element Method Analysis of the Spreading Mechanism and Its Influence on Powder Bed Characteristics in Additive Manufacturing. *Micromachines*, 12, 392. https://doi.org/10.3390/mi12040392

[3] Chen, H., Wei, Q., Zhang, Y., Chen, F., Shi, Y., Yan, W. (2019). Powder-spreading mechanisms in powder-bed-based additive manufacturing: Experiments and computational modeling. *Acta Materialia*, 179, 158–171. https://doi.org/10.1016/j.actamat.2019.08.030

[4] Meier, C., Weissbach, R., Weinberg, J., Wall, W.A., Hart, A.J. (2019). Critical influences of particle size and adhesion on the powder layer uniformity in metal additive

manufacturing. *Journal of Materials Processing Technology*, 266, 484–501. https://doi.org/10.1016/j.jmatprotec.2018.10.037

[5] Parteli, E.J.R., Pöschel, T. (2016). Particle-based simulation of powder application in additive manufacturing. *Powder Technology*, 288, 96–102. https://doi.org/10.1016/j.powtec.2015.10.035

[6] Srivastava S., Garg R.K., Sharma V.S., Alba-Baena N.G., Sachdeva A., Chand R., Singh S. (2020) Multi-physics continuum modelling approaches for metal powder additive manufacturing: a review. *Rapid Prototyping Journal, 26 (4), pp. 737 – 764 DOI: 10.1108/RPJ-07-2019-0189.*

[7] Singh A., Kapil S., Das M. (2021) Discrete Element Analysis of Gravity-Driven Powder Flow in Coaxial Nozzles for Directed Energy Deposition. *Springer Proceedings in Materials, 9, pp. 313 – 332 DOI: 10.1007/978-981-16-0182-8_24.*

[8] Chuchuay P., Khemabulkul K., Ninpetch P., Kowitwarangkul P. (2024) Selective Laser Melting of Titanium Alloys: Simulation of Scanning Speed Effects with High Layer Thickness. *Materials Science Forum, 1141, pp. 11 – 18 DOI: 10.4028/p-W5xr6A.*

[9] Zohdi T.I. (2018) DEM extensions: Electrically driven deposition of polydisperse particulate powder mixtures. *Lecture Notes in Applied and Computational Mechanics, 60, pp. 121 – 134 DOI: 10.1007/978-3-319-70079-3_7.*

[10] Bouabbou A., Vaudreuil S. (2022) Understanding laser-metal interaction in selective laser melting additive manufacturing through numerical modelling and simulation: a review. *Virtual and Physical Prototyping, 17 (3), pp. 543 – 562 DOI: 10.1080/17452759.2022.2052488.*

[11] Abdelkrim B., Vaudreuil S. (2024) An Open-Source Discrete Element Model for SS316L Alloy Powder Characterization Using a Virtual Hall-Flow Meter to Study the Flowability in Powder Bed Fusion Additive Manufacturing. *Springer Tracts in Additive Manufacturing, Part F3253, pp. 151 – 159 DOI: 10.1007/978-3-031-32927-2_14.*

[12] Tong M. (2023) Review of Particle-Based Computational Methods and Their Application in the Computational Modelling of Welding, Casting and Additive Manufacturing Metals, 13 (8), art. no. 1392 DOI: 10.3390/met13081392.